\documentclass[prd,preprint,eqsecnum,nofootinbib,amsmath,amssymb,
               tightenlines]{revtex4}
\usepackage{graphicx,color}
\usepackage{bm}
\def\be{\begin{eqnarray}}
\def\ee{\end{eqnarray}}
\def\t{{\hat t}}
\def\z{{\hat z}}
\def\y{{\hat y}}
\def\Y{{\hat Y}}
\def\u{{\hat u}}
\def\U{{\hat U}}
\def\V{{\hat V}}
\def\F{{\hat F}}
\def\w{{\hat \omega}}

\def\v{{\bf v}}
\def\x{{\bf x}}
\def\n{{\hat n}}

\def\llangle{\left\langle}
\def\rrangle{\right\rangle}

\def\Eq#1{Eq.~(\ref{#1})}
\def\st{\begin{equation}}
\def\stp{\end{equation}}
\def\bg{\begin{eqnarray}}
\def\nd{\end{eqnarray}}

\def\Fig#1{Fig.~\ref{#1}}

\def\Ref#1{Ref.~\cite{#1}}
\def\Refs#1{Refs.~\cite{#1}}

\def\AdS{\mbox{AdS}}

\def\N{\mathcal{N}}
\def\r{{\bf r}}
\def\Tr{{\rm Tr}}
\def\tr{{\rm tr}}
\def\k{{\bf k}}
\def\A{{{A}}}
\def\zero{{\underline{0}}}
\def\one{{\underline{1}}}
\def\two{{\underline{2}}}

\def\bra#1{{\left\langle{#1}\right|\,}}
\def\ket#1{{\,\left|{#1}\right\rangle}}
\def\C{{\scriptscriptstyle C}}
\def\tt{{t}}

\begin{document}

\vspace*{1cm}

\title{Transverse Momentum Broadening of a Fast Quark in a $\N=4$ Yang Mills Plasma}
\author{Jorge Casalderrey-Solana}
\affiliation
    {%
    Nuclear Science Division, MS 70R319, Lawrence Berkeley National Laboratory, Berkeley, CA 94720
    }%
\author{Derek Teaney}
\affiliation
    {%
    Department of Chemistry \& Physics, Arkansas State University, State University, Arkansas 72467, USA
    }%
\date{\today}

\preprint {LBNL-62161}

\begin{abstract}
We compute the momentum broadening of a heavy fundamental charge
 propagating  through a 
$\mathcal{N}=4$ Yang Mills plasma  at large 
t' Hooft coupling. We do this by expressing the medium modification
of the probe's density matrix in terms of a Wilson loop 
averaged over the plasma. 
We then use the AdS/CFT correspondence to
evaluate this loop, by identifying the dual semi-classical string solution. 
The calculation introduces the  
type ``1'' and type ``2'' fields of the thermal field theory 
and associates the corresponding sources 
with the two boundaries of the AdS space containing a black hole.
The transverse fluctuations 
of the endpoints of the string determine  
$\kappa_T = \sqrt{\gamma \lambda} T^3 \pi$ 
-- the mean squared momentum transfer per unit time.
($\gamma$ is the Lorentz gamma factor of the quark.)  
The result reproduces previous results for the diffusion
coefficient of a heavy quark. We compare our results with
previous AdS/CFT calculations of $\hat{q}$.
\end{abstract}
\maketitle

\section{Introduction}
The suppression of high transverse momentum particles in 
nucleus-nucleus collisions known as jet quenching 
can be used to calibrate  
the properties of the hot and dense matter
formed at RHIC \cite{Adcox:2001jp,Adler:2002xw}. 
This suppression is the result of the energy loss
of partons propagating through the medium. 
For sufficiently high transverse momentum,  radiation is expected  to
be the dominant energy loss mechanism.

There are several approaches to computing the radiative energy loss 
\cite{Gyulassy_losses,Kovner:2003zj,Dok_etal}. 
Here we will focus 
on the multiple soft scattering approximation, known as the
BDMPS approach after its authors \cite{Kovner:2003zj,Dok_etal}. In this 
formalism, the medium induced radiation is a function of the transport
coefficient $\hat{q}=2\kappa_T$, which is the mean squared momentum transfer
per unit time
that is imparted to the gluon as it
traverses the medium\footnote{The more common definition is the mean squared momentum transfer per unit length rather than  time.  The two definitions coincide in the ultra-relativistic limit.}.
The value of $\hat{q}$ was
estimated from the suppression of high $p_T$ particles observed at RHIC,\, 
   $\hat{q}\approx 10-15\, \rm{GeV^2/fm}$ \cite{Eskola:2004cr,Dainese:2004te}.
This value is surprisingly large when compared to the 
typical momentum scale of the medium $T\sim 300 $ MeV.

The strong jet quenching 
and the  
strong elliptic flow
\cite{Adler:2003kt,Ackermann:2000tr,Kolb:2001qz,Teaney:2001av}
suggest that the matter is strongly interacting.
For this reason, 
it is useful to have a foil to weak 
coupling  calculations of transport in high temperature gauge 
theories.
This foil is provided by 
$\mathcal{N}=4$ Super Yang Mills (SYM) at large t' Hooft coupling,
which is tractable  once the 
AdS/CFT conjecture is accepted \cite{Maldacena:1997re,Gubser:1998bc,Witten:1998qj}.
This conjecture states that $\mathcal{N}=4$  SYM is dual to Type II B
string theory in an $\AdS_5\times S_5$ background. The correspondence is
particularly attractive because real time correlators in  the strong coupling limit of the gauge
theory  can be determined by solving classical super gravity equations of motion.

In the field of heavy ion phenomenology, interest in the AdS/CFT correspondence  began when Policastro, Son and Starinets 
determined the shear viscosity ($\eta$) to entropy ($s$)  ratio
of the $\N=4$ SYM plasma  \cite{Policastro:2001yc} 
\st
         \frac{\eta}{s}  = \frac{1}{4 \pi} \, .
\stp
This ratio says that the mean free path is a fraction 
of the inverse temperature, $\ell_{mfp} \sim 1/(4\pi T) $. 
Since an $\eta/s$ ratio of this order is necessary
to have hydrodynamics at RHIC, the computation in $\N=4$
SYM was important, because it showed that $\eta/s$ can be this
small, at least in some  specific theories.


Since then there has been a lot of activity in this field. 
The drag coefficient of a heavy quark was computed in Refs.~\cite{Yaffe,Gubser:2006bz} and agrees with the diffusion coefficient
found in Ref.~\cite{Casalderrey-Solana:2006rq}. 
The drag also has been  computed  in different backgrounds
\cite{Herzog:2006se,Caceres:2006dj,Sin:2006yz,Chernicoff:2006yp,Talavera:2006tj}.
The properties of bound states of heavy quarks 
was studied in Refs.~\cite{Liu:2006nn,Peeters:2006iu,Chernicoff:2006hi,Friess:2006rk,Argyres:2006vs,Avramis:2006em}, and
the fields associated with the jet's passage were studied in 
Refs.~\cite{Friess:2006fk,Friess:2006aw}.
In addition, the conjecture has been used to describe the initial collision \cite{Shuryak:2005ia,Lin:2006rf},
and the subsequent hydrodynamic evolution \cite{Janik:2006gp,Friess:2006kw,Sin:2006pv}.  

In \Refs{Liu:2006ug,Liu:2006he}, the ``jet-quenching parameter''  was
computed by taking the dipole formula \cite{Zakharov:1997uu}
as a non-perturbative definition of $\hat{q}$ \cite{Kovner:2003zj}. The computation proceeds by 
evaluating a light like Wilson loop running along the lightcone. 
This computation has been extended in different ways by several 
authors \cite{Buchel:2006bv,Vazquez-Poritz:2006ba,Avramis:2006ip,Armesto:2006zv}.

In this paper we will define $\kappa_T$ as the mean
squared transverse momentum transfer to a heavy 
quark propagating through the medium \cite{Baier:1996sk}. Along the lines 
of \Ref{Casalderrey-Solana:2006rq}, we compute the medium modifications 
of the heavy quark density matrix, which in turn  
is related to the momentum broadening.\footnote{
The momentum broadening can be expressed as a Wilson loop which is 
similar to that of \Refs{Liu:2006ug,Liu:2006he}, but with 
the Wilson lines approaching the lightcone from below.}
 The (1,2) structure of the density matrix is 
identified with the two boundaries of the Kruskal plane of the AdS
black hole.  The computation proceeds by fluctuating an appropriately
defined Wilson line. A careful analysis of the fluctuations leads to 
\be
\kappa_T=\sqrt{\gamma \lambda} T^3 \pi \, ,
\ee
which diverges in the ultrarelativistic limit.
This value is different from that of Refs.~\cite{Liu:2006ug,Liu:2006he}, 
which is numerically close to the zero velocity limit
of this expression. An analysis of the approximations underlying 
the dipole formula used in these works, and the relationship of $\hat{q}$ to the squared
momentum transfer, may shed light on the discrepancy, and  
clarify the dynamics of the strongly coupled medium.

\section{Transverse Momentum Broadening}
We will for simplicity consider a scalar heavy ``quark'' coupled to
gauge fields.
The  heavy quark is described by the
heavy quark effective Lagrangian 
\st
\label{Lagrange}
 \mathcal{L} = \mathcal{L}_{YM} + Q^\dagger\left(iu\cdot D - M\right) Q \, .
\stp
Here we follow the conventions, 
$u^{\mu} = (\gamma, \gamma {\bf v})$, $v^{\mu}=(1,\v)$,
$D_{\mu} = \partial_{\mu} + i A_{\mu}$, and the link from the origin to ${\rm d}X$ is $U({\rm d}X, 0) = e^{-i {\rm d} X^{\mu} A_{\mu} }$ 

Following the formulation of QCD kinetic theory 
\cite{Baym,Blaizot:2001nr}, we define the Wigner distribution
\st
f_{cd}(X, \r_\perp)  = 
U_{ca}(X, X + \r_\perp/2)\; Q_{a} (X + \r_\perp/2) \rho  
Q^{\dagger}_{b} (X - \r_\perp/2)\; U_{bd}(X-\r_\perp/2, X)  \, ,
\stp
where $X$ denotes the four vector, $X=(t,\x)$, $\r_\perp$ 
is a transverse displacement, and $X+\r_\perp$ denotes (with 
a small abuse of notation) the space time point,  $(t, \x_\perp + \r_\perp, z)$.
Then the color indices are traced and  the Wigner function is 
averaged with the density matrix of the gauge + quark ensemble 
\st
f(X, \r_\perp) \equiv 
\llangle f_{cc} (X,\r_\perp) \rrangle 
= \Tr\left[ \rho \; Q^{\dagger}_{a} (X - \r_\perp/2) U_{ab} Q_b(X+\r_\perp/2)\right] \, , 
\stp
with $U_{ab}$ the straight link 
$U_{ab}(X-\r_\perp/2, X + \r_\perp/2)$ .

To motivate this definition and subsequent developments, we note that in kinetic theory the Fourier transform of this object is identified
with the phase space distribution
\st
 f(X,\k_\perp)  = \int d^2\r_{\perp}\,  e^{-i\k_{\perp} \cdot \r_{\perp}} \, f(X, \r_\perp) \, .
\stp
In this way the transverse current is 
\st
  \int \frac{d^2\k_\perp}{(2\pi)^2}\, \k_{\perp} f(X,\k_\perp) = 
-i\frac{\partial}{\partial \r_\perp}  f(X, \r_\perp)  = \Tr \left[\rho \;  \left\{ Q^{\dagger} \left(-\frac{i}{2} {\bf D}_{\perp} Q \right) + \left({-\frac{i}{2} \, \bf D_\perp} Q\right)^\dagger Q\right\}  \right] \, .
\stp
With this introduction, 
we will identify  the mean transverse momentum squared with
\bg
  \llangle {\bf p}_{\perp}^2(t) \rrangle &=& \int d^3\x\, \int \frac{d^2\k_\perp}{(2\pi)^2}\, \k_{\perp}^2 f(X,\k_\perp)\, , \\ &=& \int d^3\x \, \left. -\nabla^2_{\r_\perp} f(X, \r_\perp) \right|_{\r_\perp = 0} \, .
\label{pt2}
\nd

If the heavy particle starts with a narrow transverse momentum 
distribution, then after 
a long time ${\mathcal T}$ 
we expect that the average squared transverse momentum  is
\bg
\label{pbroad}
 \llangle {\bf p}_{\perp}^2({\mathcal T})\rrangle =  2\kappa_{T} \mathcal T \, ,
\nd
where the factor of two accounts for the two transverse directions, 
and $\kappa_{T}$ is the momentum diffusion coefficient.

We next study the time evolution of the Wigner function.
A complete set of states with one heavy quark may be written
with the short hand notation
\st
\label{set_of_states}
   \sum_{\A_1} \int_{\one} \,  Q^{\dagger}(\underline{1}) 
\ket{\A_1} \bra{\A_1}  Q(\underline{1}) 
\equiv 
   \sum_{A_{\mu}, a_1} \int d^3\x_1 \,  Q^{\dagger}_{a_1} (\x_1) 
\ket{A_{\mu}} \bra{A_{\mu}}  Q_{a_1}(\x_1)  \, ,
\stp
where  $\A_{1}$ labels the Eigenstates of the gauge 
fixed operator $A_{\mu}(\x)$ (see \Ref{Weinberg} for 
a convincing discussion of constraints). 
Then, working in the Schr\"odinger picture, we first
evaluate the density matrix
\st
\Tr[\rho(t)]  = \Tr\left[ e^{-i H (t-t_o)}\, \rho(t_o) \, e^{+i H (t- t_o)} \, \right] \, ,
\stp 
were $\rho(t_o)$ is the density matrix at time $t_o$.
Inserting complete set of states  
we obtain 
\bg
\Tr[\rho(t)] & = & { \sum_{{ \A_{0}\A_{1}\A_{2}}}}\int_{\zero, \one,\two} \, \bra{\A_0}Q(\zero)  \,e^{-i H(t-t_o)}\, Q^{\dagger} (\one) \ket{\A_1} \nonumber \\
& \times & \bra{\A_1} Q(\one)  \rho(t_o)  Q^{\dagger}(\two) \ket{\A_2} \nonumber \\
& \times & \bra{\A_2} Q(\two) e^{+i H (t-t_o)} Q^{\dagger}(\zero) \ket{\A_0} \, . 
\nd
Putting the first term at the end,  we rewrite this 
as the path integral
\st
   \int_{\x_1\x_2} \int [DA_{\mu}] [ DQ DQ^\dagger]  \,
  \rho^{o}_{a_1a_2}[\x_1, \x_2, \A_1, \A_2]\, e^{i \int_{\scriptscriptstyle C}\, d^4x_c \, {\mathcal L}_{YM} +  Q^{\dagger} i u\cdot D  - M Q } 
\, Q_{a_2} (\x_2, t_o - i\epsilon) \; 
 Q^{\dagger}_{a_1} (\x_1, t_o) \, ,
\stp
with 
\st
   \rho_{a_1 a_2}^{o}[\x_1,\x_2,\A_1, \A_2]  =  \bra{\A_1} Q_{a_1}(\x_1) \rho(t_o) Q_{a_2}^{\dagger}(\x_2) \ket{\A_2} \, .
\stp
Here the path integral is performed along the closed
time path, starting from $t_o$, running up to time $t$, and
returning back to $t_o-i\epsilon$.

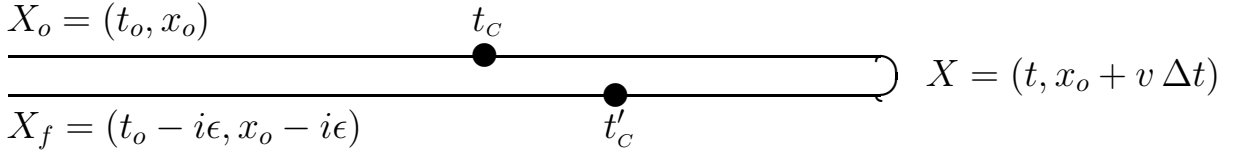
\begin{figure}
\setlength{\unitlength}{3947sp}%
\begingroup\makeatletter\ifx\SetFigFont\undefined%
\gdef\SetFigFont#1#2#3#4#5{%
  \reset@font\fontsize{#1}{#2pt}%
  \fontfamily{#3}\fontseries{#4}\fontshape{#5}%
  \selectfont}%
\fi\endgroup%
\begin{picture}(5790,906)(-200,-1774)
\put(1726,-1036){\makebox(0,0)[lb]{\smash{{\SetFigFont{14}{16.8}{\rmdefault}{\mddefault}{\updefault}{\color[rgb]{0,0,0}$t_{\scriptscriptstyle C}$}%
}}}}
{\color[rgb]{0,0,0}\thinlines
\put(1801,-1186){\circle*{150}}
}%
{\color[rgb]{0,0,0}\put(2626,-1441){\circle*{150}}
}%
\thicklines
{\color[rgb]{0,0,0}\put(-1184,-1198){\line( 1, 0){5447}}
}%
{\color[rgb]{0,0,0}\put(4263,-1444){\line(-1, 0){5447}}
}%
\put(4573,-1383){\makebox(0,0)[lb]{\smash{{\SetFigFont{14}{16.8}{\rmdefault}{\mddefault}{\updefault}{\color[rgb]{0,0,0}$X= (t, x_o + v\,\Delta t)$}%
}}}}
\put(-1199,-1711){\makebox(0,0)[lb]{\smash{{\SetFigFont{14}{16.8}{\rmdefault}{\mddefault}{\updefault}{\color[rgb]{0,0,0}$X_f=(t_o - i\epsilon , x_o - i\epsilon)$}%
}}}}
\put(-1199,-1036){\makebox(0,0)[lb]{\smash{{\SetFigFont{14}{16.8}{\rmdefault}{\mddefault}{\updefault}{\color[rgb]{0,0,0}$X_o=(t_o , x_o)$}%
}}}}
\put(2551,-1711){\makebox(0,0)[lb]{\smash{{\SetFigFont{14}{16.8}{\rmdefault}{\mddefault}{\updefault}{\color[rgb]{0,0,0}$t'_{\scriptscriptstyle C}$}%
}}}}
{\color[rgb]{0,0,0}\put(4265,-1444){\oval(  4,  0)[bl]}
\put(4265,-1321){\oval(246,246)[br]}
\put(4265,-1321){\oval(246,246)[tr]}
\put(4265,-1198){\oval(  4,  0)[tl]}
}%
\end{picture}%
\caption{The contour world line of the heavy quark with 
velocity $v$ in the $z$ direction.  
The world line of the quark passing through $X_o$ 
may be parametrized 
as $X^{\mu}_{\scriptscriptstyle X_o} (\tt_{\C}) = X_o^{\mu} + v^{\mu}(\tt_\C - \tt_{o\,\C})$, with $v^{\mu}=(1,\v)$.  The Wilson line $W_{\C}[0]$ follows the
world line of the quark. The circles at $\tt_\C$ and $\tt'_{\C}$ 
indicate the insertions of the field strengths $F^{y\mu}v_{\mu}(\tt_\C)$
and $F^{y\nu}v_{\nu}(\tt'_\C)$ into the Wilson line as in
\Eq{fstrengths} .
}
\label{path}
\end{figure}

The energetic quark may be integrated out of this expression
for the density matrix.
The contour Green Function of the heavy quark field 
in a fixed gauge background  is written
\st
   i G(2,1)  =   \llangle T_{\C} \,  Q_{a_2}(\x_2t_{2\C}) 
Q^\dagger_{a_1} (\x_1 t_{1\C})   \rrangle \, ,
\stp
where $a_i$ denote color indices, 
$t_i\C$  is the contour time, and $T_{\C}$ denotes the 
contour ordered product. This Green function satisfies
\st
\label{GF}
 (i u\cdot D - M) \, iG(2,1)
= i\delta_{a_2 a_1}  \delta^3
(\x_2-\x_1)\delta_{C}(t_{2\C}-t_{1\C})
\, ,
\stp
which has solution
\bg
 iG(2,1) &=&
e^{+iM u\cdot(X_2 - X_1)} \int_{C} \frac{d\tt_\C}{\gamma} \,
\theta(\tt_\C-\tt_{1\C}) 
\, \delta^4_{C}(X_2  - X_{\scriptscriptstyle X_1}(\tt_\C)) \qquad\qquad\qquad\qquad\,  \nonumber \\
 & &\qquad\qquad \qquad \qquad \quad \times \left[P\exp\left(-i \int_{\tt_{1\C}}^{\tt_\C} 
d\tt'_\C\,  v^{\mu} A_{\mu}(X_{\scriptscriptstyle X_1}(\tt'_\C)) \right) \right]_{a_2a_1} \, ,
\nd
where $X_{\scriptscriptstyle X_1}^{\mu}(\tt_\C)  = X_1^{\mu} + v^{\mu} (\tt_\C - \tt_{1\C})$ is the heavy quark world line which passes through $X_1$.
Then integrating the quark fields we have
\st
\label{trrho}
  \Tr[\rho(t)] = 
   \int_{\x_o} \int [DA_{\mu}] \,
   e^{i \int_{\scriptscriptstyle C}\, d^4x_c \, {\mathcal L}_{YM}  }
\,  \det\left(iu\cdot D - M\right) \, 
   \rho^{o}_{a_1a_2}[\x_o, \x_o, \A_1, \A_2]\, 
     W_C[0]_{a_1a_2} \, ,
\stp
where  
\st
    W_{C}[0] = T_{\C} \exp\left(-i\int_C d\tt_{\C} \, v^{\mu} A_{\mu}\left(X_{\scriptscriptstyle X_o}(\tt_\C) \right)\right) .
\stp
The path of the Wilson line runs along the 
world line of the heavy quark and returns on the 
time reversed path as shown in \Fig{path} .

We make three comments:
(i) Up terms suppressed by powers of $T/M$ the determinant may be 
dropped in this expression. This determinant is responsible for 
effective vertices which appear in the light quark 
Lagrangian when the heavy quark is integrated out \cite{Braaten:2000qz}.
(ii) Due to the overall translational invariance of the problem,
the average is independent of $\x_o$, and yields an overall 
factor of volume. 
(iii) We will denote the contour path integral over the 
gauge fields in \Eq{trrho}  as $\llangle\dots \rrangle_{A}$,
with the initial and final values of the gauge fields given by 
$A_1$ and $A_2$ respectively. $A_1$ and $A_2$ are
integrated over with weights given by the initial
density matrix.
Thus \Eq{trrho} becomes
\st
\label{trrho2}
    \Tr[\rho(t)] = V\, \llangle \tr\,  \rho^{o}[\x_o,\x_o,A_1, A_2] \ 
W_C[0]  \rrangle_{A} \, ,
\stp
where the  trace is over the color indices.

Next we consider the evolution  of  $f(X, \r_\perp)$
\st
f(X, \r_\perp)  = \Tr\left[ e^{-i H (t-t_o)}\, \rho(t_o) \, e^{+i H (t- t_o)} \, Q^{\dagger}_{a} (X - \r_\perp/2) U_{ab} Q_b(X + \r_\perp/2)\right] \, .
\stp 
Inserting complete set of states  as before 
we obtain 
\bg
f(X, \r_\perp) & = & { \sum_{{ \A_{0}\A_{1}\A_{2}\A_{3}}}}\int_{\zero, \one,\two} \, \bra{\A_0}Q(\zero)  \,e^{-i H(t-t_o)}\, Q^{\dagger} (\one) \ket{\A_1}  \nonumber \\
& \times & \bra{\A_1} Q(\one)  \rho(t_o)  Q^{\dagger}(\two) \ket{\A_2}\nonumber \\
& \times & \bra{\A_2} Q(\two) e^{+i H (t-t_o)} \;  
 Q^{\dagger}_a(X - \r_\perp/2) U_{ab}\ket{\A_3} \nonumber \\
& \times & \bra{\A_3}  Q_b(X + \r_\perp/2) Q^{\dagger}(\zero) \ket{\A_0} \, . 
\nd
Putting the first term at the end,  we rewrite this 
as the path integral
\bg
& &   \int_{\x_1\x_2} \int [DA_{\mu}] [ DQ DQ^\dagger]  \,
  \rho_{a_1a_2}^o[\x_1, \x_2, \A_1, \A_2]\, e^{i \int_{\scriptscriptstyle C}\, d^4x_c \, {\mathcal L}_{YM} +  Q^{\dagger} \left( i u\cdot D - M\right) Q } \, \nonumber \\
& &\;\; \times  \, Q_{a_2}(\x_2, t_o - i\epsilon) \;  Q^{\dagger}_{a}(X-\r_\perp/2) U_{ab} Q_b(X + \r_\perp/2) \; Q^{\dagger}_{a_1} (\x_1, t_o)  \, .
\nd

\begin{figure}
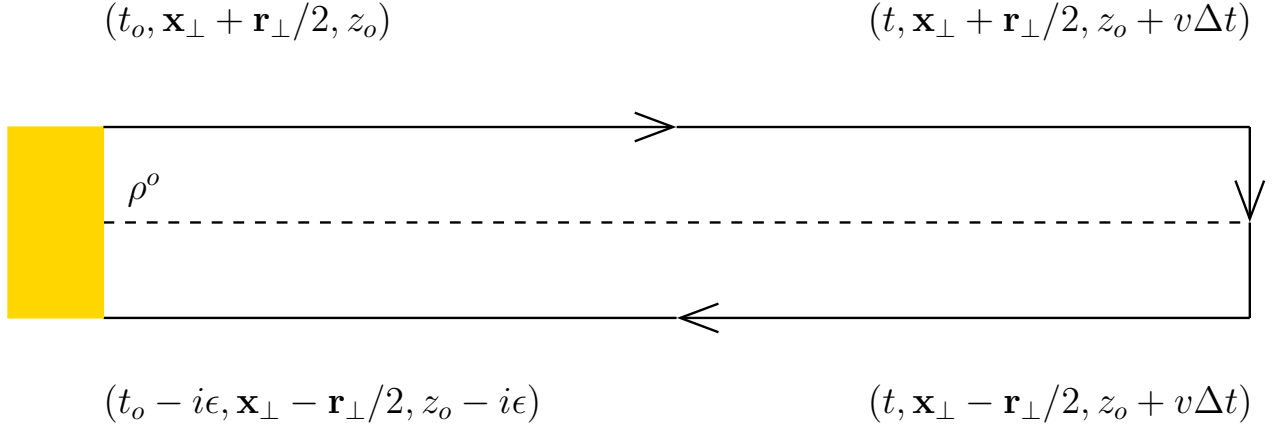
\caption{
Graphical representation of \Eq{frperp}.  The Wilson
line indicated by the black line is denoted $W_{\C}[\r_\perp/2, -\r_\perp/2]$. This Wilson line is traced with the initial 
density matrix, $\rho^{o}_{a_1 a_2}$.
}
\label{path2}
\end{figure}

Performing the quark integration,  we have
\st
\label{frperp}
    f(X, \r_\perp) = \big\langle \tr \, \rho^o[\x_o + \r_\perp/2,\x_o -\r_\perp/2,A_1, A_2] \, W_C[\r_\perp/2, -\r_\perp/2] \big \rangle_{A} \, ,
\stp
where the Wilson line $W_{\C}[\r_\perp/2, -\r_\perp/2]$ is
shown by the black lines in \Fig{path2}.
We may expand this for small $\r_\perp$  by
inserting unity
$U(X_o+\r_\perp/2, X_o) U(X_o, X_o+\r_\perp/2)$ and 
$U(X_o-\r_\perp/2, X_o) U(X_o, X_o-\r_\perp/2)$ 
at the beginning and ends of the contour  and using the 
definition of $f_{ab}$. We have
\bg
\label{broad}
   f(X, \r_\perp) &=&  f(X, 0) +   
\frac{\r_\perp^2}{2} \llangle 
\frac{\partial^2}{\partial \r^2_\perp} f_{ab} (X_o,\r_\perp) W_{C}[0]_{ab}   
\rrangle_{A}  + \mbox{momentum broadening}\, , \qquad 
\nd
with the momentum broadening given  by
\bg
\frac{1}{2} \left(\frac{\r_\perp}{2}\right)^2 \int_{\C}\int_{\C} d\tt_\C d\tt_\C'&&   
 \bigg\langle \tr \rho^{o} [\x_o,\x_o, A_1, A_2] \;
   T_{\C}\left[ F^{y \mu}(\tt_\C) v_{\mu} F^{y\nu}(\tt'_\C) v_{\nu}\,  W_{C}[0]\,  \right] \bigg\rangle_{A} \, . \qquad
\label{fstrengths}
\nd
We again make three comments: (i)  The  first term is in \Eq{broad}
is characteristic of the initial momentum  distribution of
the quark  which may be supposed small; the momentum 
broadening is described by the second term. 
(ii) 
We have tacitly assumed that $\Tr[\rho(t)]$ was unity, 
we will divide by \Eq{trrho2} to set the normalization. 
(iii) The $T_\C$ 
ordered product of field strengths means that we insert field
operators into
the Wilson line $W_C[0]$ as shown in \Fig{path}. 
As in \Ref{Casalderrey-Solana:2006rq},  we define the contour Wilson line $W_{C}[\delta y]$,
with deformations in the $y$ direction at various 
points along the contour, $\delta y(\tt_\C)$.
Then, the contour ordered product of field strengths
can be written as the variation of $W_{\C}[\delta y]$ at
times $\tt_\C$ and $\tt'_\C$ respectively.  
The shape of the Wilson line is 
the source for the contour ordered product of fields.

Thus from \Eq{pt2}, \Eq{pbroad}, and \Eq{fstrengths},  the momentum broadening is 
\st
\kappa_{T} {\mathcal T}  = \frac{1}{4} \, \frac{1}{\llangle  \tr \rho^o W_C[0] \rrangle_{A}} 
\int_\C \int_C d\tt_\C d\tt'_\C \, \llangle \tr \, \rho^o[\x_o,\x_o, A_1, A_2] \,
\frac{ \delta^2 W_{C}[\delta y] }{\delta y(\tt_\C) \,\delta y(\tt'_\C)}  \rrangle_{A} \, ,
\stp
where $\mathcal T$ is the total real time of the process.

Now we  brake up the contour integration  into
type $``1"$ and type $``2"$ fields;
we write $\delta y_1(t)$ 
and $\delta y_2(t')$  for variations on the 
one and two branches and  use $\tt,\tt'$  
for the real part of $\tt_\C,\tt'_\C$. Then the contour
correlations are 
\bg
iG_{11}(t, t') &=& 
\frac{1}{\llangle  \tr \rho^o W_C[0,0] \rrangle_{A}} 
\, \llangle \tr \, \rho^o  \; 
\frac{ \delta^2 W_{C}[\delta y_1, 0] }{\delta y_1(t) \,\delta y_1(t')}  \rrangle_{A} \, ,  \\
iG_{22}(t, t') &=& 
\frac{1}{\llangle  \tr \rho^o W_C[0,0] \rrangle_{A}} 
\, \llangle \tr \, \rho^o  \; 
\frac{ \delta^2 W_{C}[0, \delta y_2] }{\delta y_1(t) \,\delta y_2(t')}  \rrangle_{A} \, , \\
iG_{12}(t, t') &=& 
\frac{1}{\llangle  \tr \rho^o W_C[0,0] \rrangle_{A}} 
\, \llangle \tr \, \rho^o  \; 
\frac{ \delta^2 W_{C}[\delta y_1, \delta y_2] }{\delta y_1(t) \,\delta y_2(t')}  \rrangle_{A} \, ,  \\
iG_{21}(t, t') &=& 
\frac{1}{\llangle  \tr \rho^o W_C[0,0] \rrangle_{A}} 
\, \llangle \tr \, \rho^o  \; 
\frac{ \delta^2 W_{C}[\delta y_2, \delta y_2] }{\delta y_1(t') \,\delta y_2(t)}  \rrangle_{A}  \, .
\nd
Finally, we may use approximate translational invariance to
write
\st
\label{hatqf}
\kappa_T=\lim_{\omega\rightarrow 0} \frac{1}{4}  \int dt e^{+i\omega t} 
        \left( i G_{11}(t,0) + iG_{22}(t,0) + i G_{12}(t,0)+i G_{21}(t,0)\right)    \, .
\stp
\section{The AdS/CFT Correspondence}
 
In the previous section we have identified the  source
for the contour ordered electric fields as the variation
of the trajectory of a Wilson line running along the 
contour  shown in  \Fig{path}. 
The  strategy to extract
the transverse momentum diffusion coefficient $\kappa_T$ 
parallels \Ref{Casalderrey-Solana:2006rq} closely. We first construct a 
semi-classical string which is the gravity dual of 
the unperturbed Wilson line running along the Schwinger-Keyldish
contour. Then we vary the endpoint of the string  in
order to obtain the appropriate contour 
correlation functions of electric fields or $\kappa_T$.

\subsection{Preliminaries}

First we recall previous work by us and others to 
establish notation \cite{Aharoni:1999ti,Yaffe,Gubser:2006bz, Casalderrey-Solana:2006rq}
\begin{itemize}
\item 
The metric $G_{MN}$  which corresponds to $\N=4$ 
Super Yang Mills at finite temperature is the 
AdS space with a black hole
\st
 ds^2 =  \frac{r^2}{R^2}\left[-f(r) dt^2  + d\x_{\parallel}^2\right] +  \frac{R^2}{f(r) r^2}\,dr^2 + R^2 d\Omega_5^2
\, ,
\stp
where $f(r) = 1 - \left(\frac{r_o}{r}\right)^4$, $R$ is the AdS radius,  and
$r_o$ is related to the Hawking temperature $\pi T R^2 = r_0$. 
We define the scaled units, $\pi T t = \bar{t}$, $\pi T x = \bar{x}$
and $\bar{r} = r/r_o$, and  define $\bar{z}= 1/\bar{r}$ , so that
the metric reads
\st
\frac{ds^2}{R^2} = -\frac{1}{\bar{z}^2}\, f(\bar{z})\, d\bar{t}^2 + \frac{1}{\bar{z}^2 }
d\bar{\x}_{\parallel}^2 +  \frac{d\bar{z}^2}{f(\bar{z})\,\bar{z}^2} + 
d\Omega_5^2 \, ,
\stp
and we distinguish the function $f$ by its argument,  
$f(\bar{z}) = 1 - \bar{z}^4$ .
Further, we 
will sometimes change variables to $u=\bar{z}^2$.
In what follows we will drop the ``bar".
\item 
The gravity dual of a
Wilson line moving along the unperturbed trajectory 
$x_3 = v t$ is a semi-classical string stretching in
$\AdS_5 \times S_5$ whose endpoint follows  the 
curve $x_3 = v t $.    
The dynamics of the semi-classical 
string  is described by  the Nambu-Goto action
\st
    S_{NG} = \frac{1}{2\pi \alpha'}\int d\tau d\sigma 
       \sqrt{-\mbox{det} h_{ab}}  \; ,
\stp
where 
\st
   h_{ab} = G_{MN} \partial_a X^M \partial_b X^N \,. 
\stp
The solution to the classical equations of motion 
is described by the map
\[
   (\tau,\sigma)\mapsto (t=\tau, \x_{\perp}=0, x_3(\tau,\sigma),\,  z=\sigma,\, \Omega_5 = \mbox{Const} ) \, ,
\]
with 
\st
\label{Yaffeetal}
   x_{3}(t,z) = v t +  \frac{v}{2} \left[ \arctan(z)  - {\rm arctanh}(z) \right]
\stp
\end{itemize} 

\subsection{The Semi-Classical String in the Kruskal plane}

\begin{figure}
\begin{center}
\includegraphics[width=7cm,angle=-90]{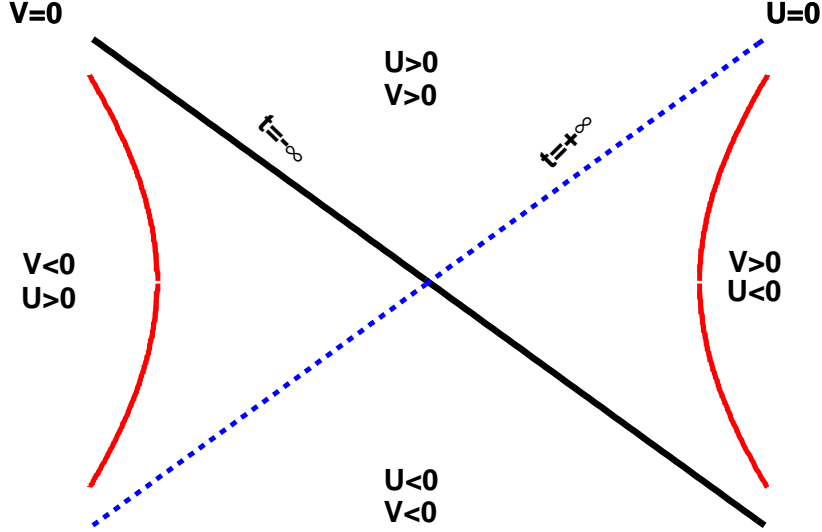}
\caption{
Kruskal diagram for the AdS black hole. The coordinates $(t,r)$ span the 
right quadrant.
The thick hyperbolas on the sides
of the two quadrants
are the
boundaries  at $r=\infty$.  The boundary 
in the right and left quadrants correspond to the 
``1" and ``2" axes of the thermal field theory respectively.
The  static quark corresponds to a  string  which
spans the full Kruskal plane.  At finite 
velocity, the asymptotic solution \Eq{Yaffeetal} 
is discontinuous at $t=-\infty$ or $V=0$.
}
\label{krusk}
\end{center}
\end{figure}

The semi-classical string solution given in \Eq{Yaffeetal} is dual to 
a quark moving with velocity $v$. In the gauge 
theory this quark is constructed by turning on
a $U(1)$ electric field and allowing the distribution
to come to stationary state \cite{Yaffe}. Far in the past the heavy quark 
is nearly at rest,
and the world sheet covers the full Kruskal 
plane \cite{Casalderrey-Solana:2006rq}\,\footnote{ 
Our Kruskal conventions are the following.
We first define 
\st
z_{*}(z) \equiv \int_0^{z} \frac{dz}{f(z) }   
         =   \frac{1}{2} \tan^{-1}(z)   +  \frac{1}{2} \tanh^{-1}(z)  
\stp
Then we define  the coordinates $(\nu_+,\nu_-)$
\st
   \nu_{+}  \equiv  t+z_*(z)  \qquad
   \nu_{-} \equiv t-z_*(z)
\stp
that for instance $\nu_{-} ={\rm Const}$ is the geodesic
of an in-falling lightlike particle. Finally the Kruskal
coordinates are 
\st
   U = -e^{-2\nu_+}  \qquad V =  e^{+2\nu_-}
\stp
We also define
$V' = V e^{2\tan^{-1}(z)}$ and $X' = - U V e^{2\tan^{-1}(z)} $.
}.
This is shown in \Fig{krusk} which also
 illustrates the Kruskal coordinates.
 As the electric field
is turned on in the gauge theory, 
the quark accelerates and slowly approaches
the stationary velocity distribution. In the gravity 
dual this corresponds to slowly turning on a $U(1)$ 
electric field in the boundary brane and waiting 
for the string to reach the asymptotic form given
by \Eq{Yaffeetal}.  We re-write this solution in Kruskal 
coordinates as
\st
\label{YaffeKrusk}
    x_3 = \frac{v}{2} \log(V) + v\,\arctan(z) \qquad \mbox{for $V>0$}\,.
\stp
Examining this solution we see a logarithmic divergence for $V\rightarrow 0$,   i.e. in the distant past,  $t \rightarrow -\infty$. This discontinuity 
in the distant past reflects the fact that the asymptotic 
solution \Eq{YaffeKrusk} is
not a solution  as the electric field 
was turned on slowly.  At $t\rightarrow -\infty$ it is reasonable expect that
this solution slowly deforms into the static solution described above
and covers the full Kruskal plane.

Therefore, we will demand analyticity across the  $V=0$ line
while extending  $V$ through the lower half plane to negative real argument\, \footnote{This choice of extending $V$ through the lower half plane is compatible
with the construction of Son and Herzog for the real time path
integral which we adopt later.}. Through this process  $x_3$ becomes imaginary
\st
    x_3 = \frac{v}{2}\log(\left|V\right|)  + v\,\arctan(z)  - i v \underbrace{\pi/2}_{\beta/2} \, .
\stp
This imaginary value for the coordinate $x_3$ is appropriate. 
The left quadrant of the Kruskal plane is associated with the 
``2'' branch of the field theory with $\sigma = i\beta/2$, or 
$i \pi/2$ for our rescaled units.
Since the operators on the real axis are evaluated 
at the space time point $(t,x_3) = (t, v t )$,  when time is extended 
in the imaginary time direction we expect that the operators
should be evaluated at   
\st
   (t, v t) \rightarrow (t - i\beta/2, v t - i v\beta/2) \, .
\stp
Thus the analytic extension of the string solution into 
the lower half $V$ plane maps to the appropriate $x_3$ 
coordinate.

In what follows  we will take  this string solution 
as the gravity dual of the heavy quark (or Wilson line) 
propagating along the Schwinger-Keyldish contour shown 
in \Fig{path}  with the second axis displaced  
by $i\beta/2$.
We will make the correspondence 
between the ``1'' and ``2'' axes of the contour and 
the endpoints of the string on the right and left boundaries of AdS. 
When studying fluctuations of this semi-classical
string solution, we will bear in mind this discussion of the 
past infinity and demand analyticity across the $V=0$ line.

\subsection{Fluctuations and World Sheet Black Hole}

Having constructed the string solution corresponding 
to a heavy quark propagating along the Schwinger Keldysh contour,
we will proceed to fluctuate the shape of the string solutions 
in the transverse directions and solve for the fluctuations.
Once the classical solution is determined in the  
Kruskal plane  we follow the general philosophy of the 
AdS/CFT correspondence and equate the classical action
of the source to the generating functional
\st
\label{generating}
     \frac{1}{e^{iS_{NG}[0,0]}} e^{iS_{NG}[\delta y_1, \delta y_2]}  
=  \frac{1}{\llangle \tr \, \rho^o\, W[0,0] \rrangle_A} \, \llangle \tr\, \rho^o\,  W[\delta y_1, \delta y_2] \rrangle_A
\, .
\stp
However in order to find the classical solution, boundary
conditions are needed.  The appropriate boundary 
conditions for $G_{11}$ and $G_{22}$  etc. are not obvious.

The appropriate boundary conditions are greatly clarified
by a change of coordinates which transforms the original
induced metric to a diagonal metric  which turns out to 
be equivalent to a world sheet black hole.
Using the original coordinates $(\tau,\sigma)=(t,z)$ the induced metric is
\be
h_{tt}&=&-\frac{R^2}{z^2}\left(
                       \frac{1}{\gamma^2}-z^4
                       \right) \, ,
\\
h_{zz}&=&\frac{R^2}{z^2}\frac{1}{f^2}
                       \left(
                       1-\frac{z^4}{\gamma^2}
                       \right) \, ,
\\
h_{tz}&=&-\frac{R^2}{z^2}\frac{1}{f}v^2z^2 \, .
\ee
This non-diagonal world sheet metric can be diagonalized by the following change 
of coordinates 
\be
\label{hatc}
\hat{t}&=&\frac{1}{\sqrt{\gamma}}
       \left(t +\frac{1}{2}\arctan(z) - 
                \frac{1}{2}\sqrt{\gamma} \arctan(\sqrt{\gamma} z)
               -\frac{1}{2} \rm{arctanh}(z)
               +\frac{1}{2}\sqrt{\gamma} \rm{artanh} (\sqrt{\gamma}z)
       \right) \, ,
\\
\hat{z}&=&\sqrt{\gamma} z \, .
\ee
and the metric takes the form 
\be
h_{\t\t}&=&-\frac{R^2}{\z^2}f(\z) \, ,\\
h_{\z\z}&=&\frac{R^2}{\z^2}\frac{1}{f(\z)} \, , \\
h_{\t\z}&=&0 \, ,
\ee
with $f(\z)=1-\z^4$. 
Thus the  world sheet metric is in fact a 
black hole with the horizon at $z^2=1/\gamma$ or $\hat{z}=1$
\footnote{
This fact has been already noted in \cite{Gubser:2006nz}
}. 
As usual, the horizon is a coordinate singularity  and 
is a consequence the coordinate transformation in 
Eq. (\ref{hatc}).
In this coordinate system, the induced metric for the 
moving string with ``hat'' variables equals the induced metric of the 
static string with ``un-hatted'' variables.

\begin{figure}
\begin{center}
\includegraphics[height=2.76in,width=2.7in,angle=-90]{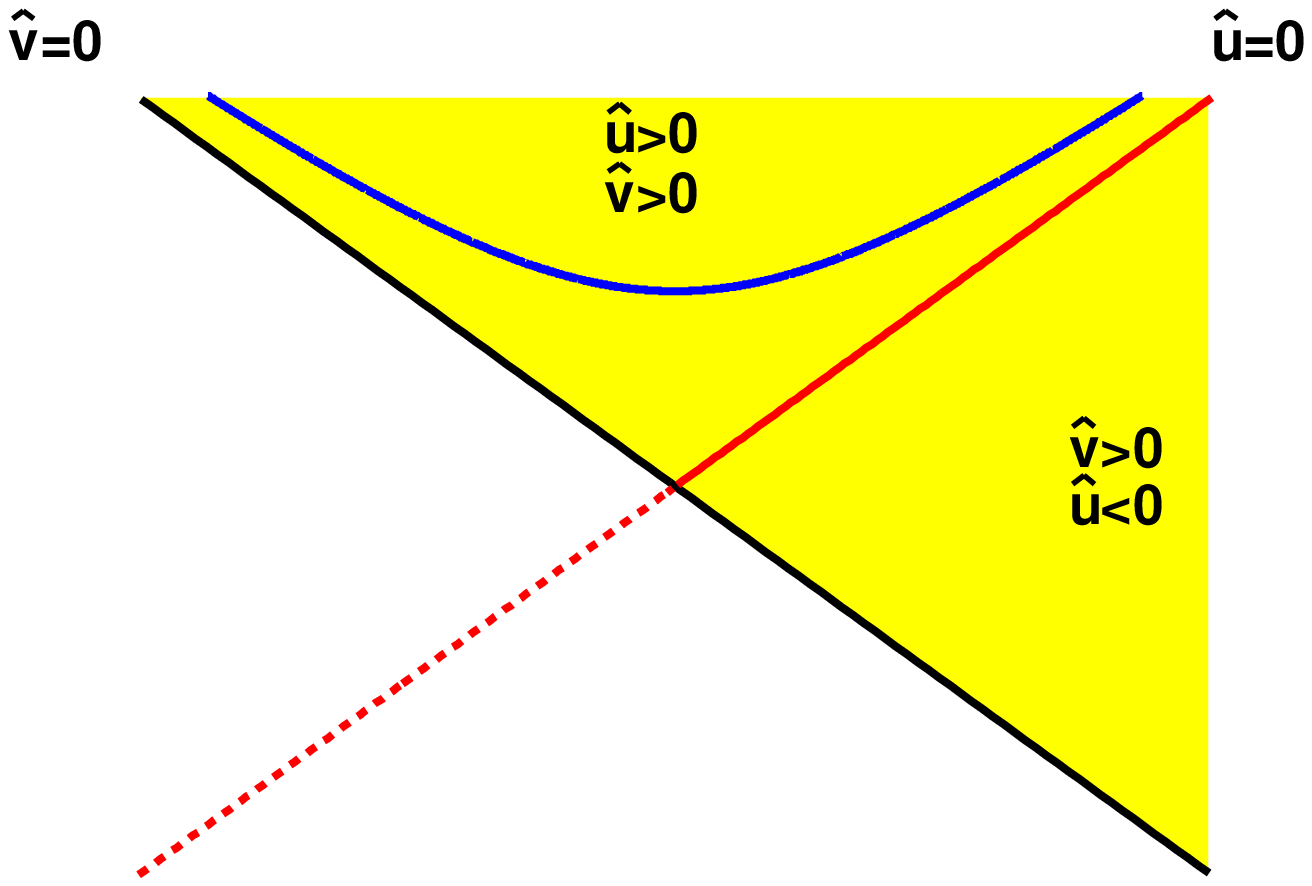}
\hspace{0.2in}
\includegraphics[height=2.7in,width=2.7in,angle=-90]{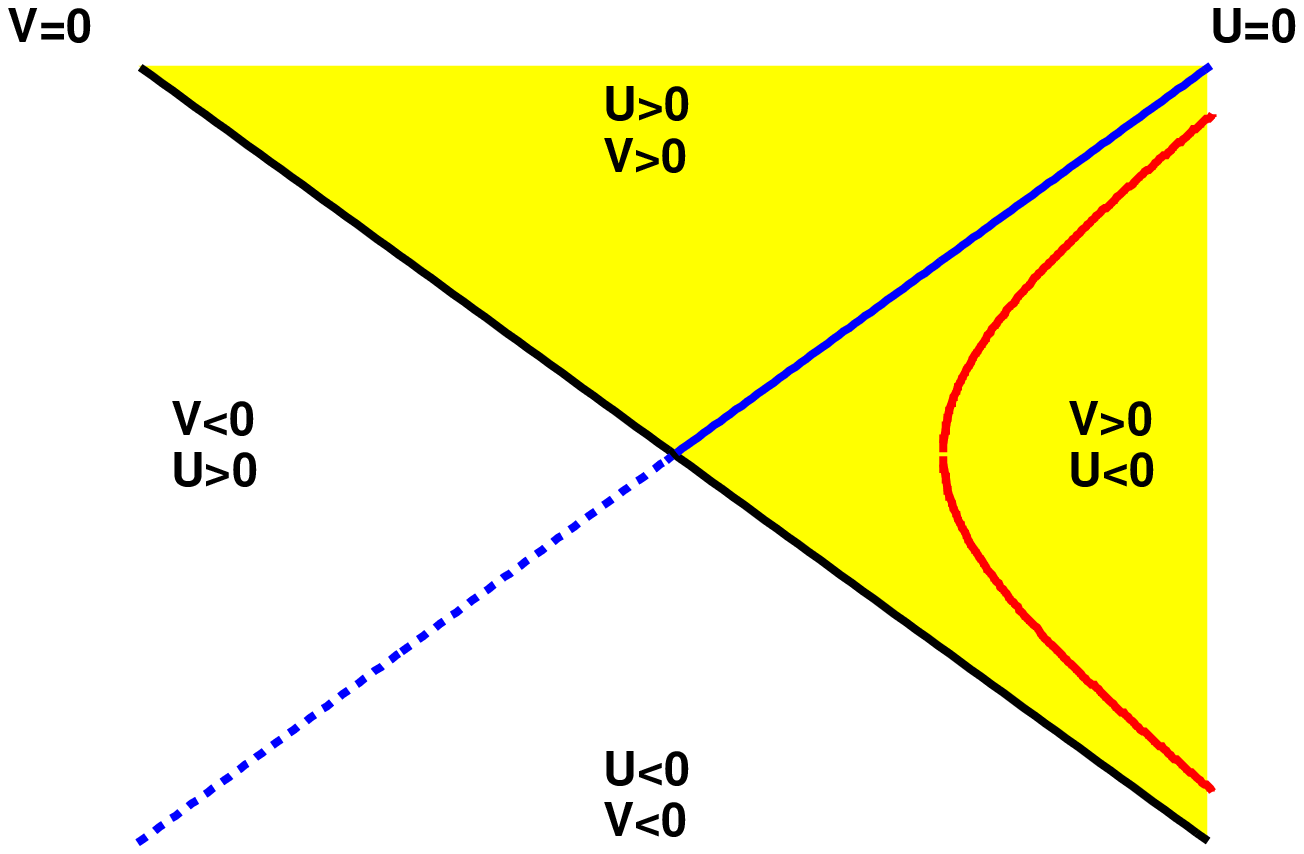}
\end{center}
\caption{
\label{world}
(a) The world sheet black hole. (b) The Schwarzschild 
black hole in space time. 
The 
future event horizon of the world sheet and its corresponding 
image $z = 1/\sqrt{\gamma}$ are indicated by the solid red line.
Similarly the solid blue line maps to the future event horizon
of the space time black hole.
The shaded region of the
world sheet black hole maps into the shaded region of 
space-time region of the Schwarzschild geometry. Similarly
there is a separate string solution in the $V<0$ half of the
$(U,V)$ plane which maps to the  $\hat{V}<0$ half of
the $(\hat{U}, \hat{V})$ plane. The two solutions 
are joined at past infinity ($V=0$) 
by demanding analyticity in the lower complex $V$ plane, i.e. 
that only positive energy solutions emerge from past infinity.
}
\end{figure}

Next we show how different regions of the world-sheet  
black hole
 map into space time. First we define the analogs
of Kruskal variables for hatted variables 
$\hat{\nu}_{-}, \hat{\nu}_+, \hat{V}, \hat{U}, \hat{z}_*,\dots $. 
(For instance 
$\hat{z}_*=(\tan^-1(\z) + \tanh^-1(\z))/2$ ). Taking $\hat{V}$ positive,
we have the following relation between the world-sheet and space-time
Kruskal variables
\footnote{
As noted before, we define
$V' = V e^{2\tan^{-1}(z)}$ and $X' = - U V e^{2\tan^{-1}(z)} $
and analogous relations for $\hat{V}'$ and $\hat{X}'$ }
\bg
   \hat{z}   &=& \sqrt{\gamma} z\, , \\
   \hat{V'}  &=& \left(V'\right)^{1/\sqrt{\gamma}} \, ,   \\
   \hat{X'}  = \frac{1-\hat{z}}{1+ \hat{z}}
 &\qquad&     X' = 
 \frac{1-{z}}{1+ {z}} \, .
\nd
From the form of this map we see the upper half
of the world sheet Kruskal plane $(\hat{U}, \hat{V})$
gets mapped into the 
upper half Kruskal plane of space-time $V>0$. 
The structure is illustrated in \Fig{world}. 
As $V$ becomes negative $\hat{V}$ becomes imaginary. This
is because the solution given in \Eq{Yaffeetal} is not valid at past infinity. In 
the lower half $(V <0)$ of the Kruskal plane  we can find a separate string
solution which maps the lower half ($\hat{V} < 0$) of
the world sheet Kruskal plane. 
At past infinity, before the electric field was turned
on, the quark is at rest, and the two string solution are
joined. We will therefore connect the solutions in the 
upper half and lower half Kruskal planes by demanding analyticity 
across the $V=0$ line, extending $V$ through the lower
half complex $V$ plane.  This extension through 
the lower complex $V$ plane is consistent with 
the Herzog-Son construction \cite{Herzog:2002pc} and physically 
says that only positive energy solutions emerge from past infinity. 
Again by analogy with the Herzog-Son construction, we 
will extend through the upper complex $\hat{U}$ plane. Physically
this says that only negative energy solutions emerge from
the  future event horizon of the world sheet black hole.

Having clarified the  analytic structure of the world
sheet black hole and determined the appropriate boundary
conditions, we next analyze small fluctuations.
The equation of motion for
small transverse fluctuations  can be found easily
in ``hatted" coordinate system.
Introducing
\be
\y=\sqrt{\gamma} y \, ,
\ee
the action for small fluctuation is the same as for 
the static string \cite{Casalderrey-Solana:2006rq}
\be
\label{action}
S_{NG}=\frac{R^2}{2\pi\alpha'}
\int \frac{d\t\,d \u}{2\u^{3/2}}
\left[
1-\frac{1}{2}\left(
             \frac{\dot{\y}^2}{f(\u)}-
              4 f(\u) \u \left(\y'\right)^2
             \right)
\right]
\, .
\ee
We define,
\be
  \y(\t,\u) = \int e^{-i \w \t}\, \y(\w)\, \Y(\w, \u)  \, \frac{d\w}{2\pi}
\, .
\ee
where we choose to normalize $\Y_\w(\u=0)=1$, {\it i.e.} 
$\y(\omega)$ is the value of the fluctuation at the boundary.

The  Euler-Lagrange equation for the small string fluctuations 
are 
\be
\partial_{\u}^2 \Y_\w - \frac{(2 + 6 \u^2)}{4 \u f} \partial_\u \Y_\w + 
\frac{\w^2}{4 \u f^2 } \Y_\w = 0
\, .
\ee
This equation is solved by 
\be
\label{wsinf}
\Y(\w,u)=\left(1-\u\right)^{- i\w/4}\F(\w, \u)
\ee
where $\F(\u)$ is a regular function of $\u>0$ .  
$(1- \u)^{-i\w/4}$ is 
 in-falling in the world-sheet horizon $\u=1$. 
The complex conjugate of this expression is also a solution of the 
differential equation and is outgoing at the horizon.

Let us now extend these solutions into 
the full Kruskal plane. It is useful 
to express the solution in terms of the world sheet 
Kruskal coordinates.
Close to $\hat{z}=1$
the in-falling solution and out-going solutions behave
as
\be
e^{-i\w \t}\Y(\w,\u)&\sim& e^{-i\w/2\ln(\hat{V})} \qquad \mbox{in-falling} \\
e^{-i\w \t}\Y^*(\w,\u)&\sim& e^{i\w/2\ln(-\hat{U})} \qquad \mbox{out-going}
\ee
As mentioned before, the coordinate $\hat{V}$ does not cross zero in this 
$z=\sqrt{u}=1/\sqrt{\gamma}$. However, $\hat{U}$  does and we should be 
careful when defining
the branch cut  for the logarithm.
In analogy with the Herzog-Son prescription, we will impose that
the solution should be analytic in the upper half of the 
complex $\U$ plane.   
Since $\U$ is negative in this quadrant, 
the prescription means that  
\be
e^{-i\w \t}\Y^*(\w,\u)\sim e^{\pi\w/2} e^{i\w/2\ln(\hat{U})} \, .
\ee
Thus, when the fluctuation crosses the world-sheet event horizon in the 
right quadrant it picks a factor $\exp\{\pi\w/2\}$. 
Due to the different relation between $\U$ and the
local $(\t,\z)$ coordinates, in the left quadrant the solution behaves as 
\be
e^{-i\w \t}\Y^*(\w,\u)\sim e^{i\w/2\ln(\hat{U})} \, .
\ee
Thus, after passing the world sheet horizon, keeping the
same analyticity properties in $\U$, the solution picks a factor
$\exp\{-\pi\w/2\}$ in the L-quadrant.

We now address how the matching of left and right quadrant is 
performed. Using the change of coordinates \Eq{hatc} we obtain
(for $\u<1$)
\be
e^{-i\w \t}\Y(\w,\u)&=&e^{-i \omega t} (1-u)^{-i\omega/4} F_i(\omega,u) \, ,\\
e^{-i\w \t}\Y^*(\w,\u)&=&e^{-i \omega t} (\frac{1}{\gamma}-u)^{i\w/2}(1-u)^{-i\omega/4} F_o(\omega,u) \, ,
\ee
where we have defined $\omega=\w/\sqrt{\gamma}$ and
$F_i(\omega,u)$, $F_o(\omega,u)$ are two regular solutions in u.
The pole at $u=1/\sqrt{\gamma}$ in the outgoing solution
is a consequence of crossing the $\U=0$ line. Note that this 
pole does not appear in the in-falling solution, since we do not
cross the $\V=0$ line.  From this
point of view, the previous prescription for the analytic properties of 
the solution in the $\U$ coordinate translates into the prescription
for going around the pole in $u=1/\sqrt{\gamma}$. Taking this prescription
into account, in the R quadrant the two solutions close to the horizon behave as
\be
\label{CHRi}
e^{-i\w \t}\Y(\w,\u)&\sim& e^{-i\omega/2 \ln(V)} \,  ,
\\
\label{CHRo}
e^{-i\w \t}\Y^*(\w,\u)&\sim& e^{\pi\w/2} e^{-i\omega/2 \ln(V)}
\, .
\ee
Note that from the point of view of the AdS black-hole both solutions
are in-falling. The exponential factor in the outgoing solution is a 
consequence of the prescription to cross the pole at $\u=1$. 
In the same way, the fluctuations in the L-quadrant behave as
\be
\label{CHLi}
e^{-i\w \t}\Y(\w,\u)&\sim& e^{-i\omega/2 \ln(-V)} \,  ,
\\
\label{CHLo}
e^{-i\w \t}\Y^*(\w,\u)&\sim& e^{-\pi\w/2} e^{-i\omega/2 \ln(-V)}
\, .
\ee

With these four expression we find four different
solutions defined in both quadrants of the (AdS) Kruskal plane
\begin{equation}
\y_{ R,i} = 
\left\{ \begin{array}{ll}
e^{-i\w \t}\Y(\w,\u) & \mbox{in R} \\
0 & \mbox{in L}
\end{array}
\right. \; \; \; \; \; \; \; \; \; \; 
\y_{ L, i} = 
\left\{ \begin{array}{ll}
0 & \mbox{in R} \\
 e^{-i\w \t}\Y(\w,\u)& \mbox{in L} 
\end{array}
\right. \ ,
\end{equation}

\begin{equation}
\y_{ R,o} = 
\left\{ \begin{array}{ll}
e^{-i\w \t}\Y^*(\w,\u) & \mbox{in R} \\
0 & \mbox{in L}
\end{array}
\right.   \; \; \; \; \; \; \; \; \; \; 
\y_{ L,o} = 
\left\{ \begin{array}{ll}
0 & \mbox{in R} \\
 e^{-i\w \t}\Y^*(\w,\u)& \mbox{in L} 
\end{array}
\right. \, .
\end{equation}

Following the Herzog-Son prescription \cite{Herzog:2002pc}, we look
for linear combinations of these expressions that,
close to the horizon, are analytic in the lower
half of the complex $V$ plane. With this criterium
only two linear combinations can be found:
\be
\y_o=\y_{R, o}+ \alpha_o \y_{L,o} \, ,\\
\y_i=\y_{R, i}+ \alpha_i \u_{L,i} \, .
\ee
From the close to horizon behaviors of the solutions 
\Eq{CHRi}, \Eq{CHRo}, \Eq{CHLi}, \Eq{CHLo}, the 
analyticity properties demand 
\be
\alpha_o&=&e^{+\pi\w}e^{ -\pi \omega/2} \, , \\
\alpha_i&=&e^{ -\pi \omega/2} \, .
\ee

These two solutions are used as a basis for the linearized 
string fluctuations defined over the full (AdS) Kruskal plane
\be
\y=\int \frac{d\w}{2\pi} \left(a(\omega)\y_o(\omega)+ b(\omega)\y_i(\omega)\right)\, .
\ee 

This prescription recovers the 
results of Herzog  and Son \cite{Herzog:2002pc} when $\gamma=1$. To see this
one has to realize that when $\gamma=1$, since
these solutions are only defined for $u<1$ we do 
not cross the pole and, thus, the exponential prefactors
in $\y_{\{L,R\}o}$ do not appear.

The coefficients $a(\w), b(\w)$ can be determined by the boundary
values of the solutions. Thus, if we have 
\footnote{
Note that at $\u=0$, $\t=t/\sqrt{\gamma}$. Thus
 $y(t,u=0)=\int\frac{d\omega}{2\pi} e^{-i\w \t}\y(\w)$, {\it i. e.}
$\y(\w)=y(\omega)$
}
\be
\y(\t,\u=0)\big|_{R}&=&\int \frac{d\w}{2\pi}e^{-i\w \t} \y_1(\w) \, ,\\ 
\y(\t,\u=0)\big|_{L}&=&\int \frac{d\w}{2\pi}e^{-i\w \t} \y_2(\w) \, ,
\ee
we obtain 
\be
\label{components}
a(\omega)&=&\frac{1}{e^{\pi\w}-1}
          \left(
           -\y_1(\w) + \y_2(\w) e^{\pi\omega/2}
          \right)  \, ,
\\
b(\omega)&=&\frac{1}{e^{\pi\w}-1 }
          \left(
          e^{\pi\w}\y_1(\w)-e^{\pi\omega/2}\y_2(\w)
          \right) \, .
\ee

We can now compute the boundary action in terms of the string solutions. 
In $(\hat{t}, \hat{u})$ coordinates
\be
S_B=\frac{R^2}{2\pi\alpha}
   \left[
   \int_R \frac{d\hat{\omega}}{2\pi} \frac{1}{\hat{u}^{1/2}}
        \y(-\hat{\omega},\hat{u})\partial_{\hat{u}} \y(\hat{\omega},\hat{u})
   -
   \int_L \frac{d\hat{\omega}}{2\pi} \frac{1}{\hat{u}^{1/2}}
        \y(-\hat{\omega},\hat{u})\partial_{\hat{u}} \y(\hat{\omega},\hat{u})
   \right]  \, .
\ee
Using  Eq. \ref{components} 
this action can be expressed as
\footnote{This expression looks slightly different that Eq. (28) of
\cite{Herzog:2002pc} because in that work $f_k(u)$ is defined as outgoing,
while here $\Y(\w,\u)$ is infalling. This notation is chosen to 
connect with our previous work \cite{Casalderrey-Solana:2006rq}.
}  
\be
S_B&=&\frac{R^2}{2\pi\alpha}\sqrt{\gamma} \int \frac{d\omega}{2\pi} 
\Big[ \\ 
&& y_1(-\omega)y_1(\omega)\left( (\n+1)\frac{1}{\hat{u}^{1/2}}
\Y^*(-\hat{\omega},\hat{u})\partial_{\hat{u}} \Y(\hat{\omega},\hat{u})
 -\n\frac{1}{\hat{u}^{1/2}}
 \Y(-\hat{\omega},\hat{u})\partial_{\hat{u}} \Y^*(\hat{\omega},\hat{u})
                         \right)   +\nonumber
\\
&&      
y_1(-\omega)y_2(\omega)e^{\pi\omega/2}\n
      \left( -\frac{1}{\hat{u}^{1/2}}
   \Y^*(-\hat{\omega},\hat{u})\partial_{\hat{u}} \Y(\hat{\omega},\hat{u}) 
+\frac{1}{\hat{u}^{1/2} }  
   \Y(-\hat{\omega},\hat{u})\partial_{\hat{u}} \Y^*(\hat{\omega},\hat{u})
      \right) +\nonumber
\\
&&
y_2(-\omega)y_1(\omega)e^{-\pi\omega/2}(1+\n)
      \left( -\frac{1}{\hat{u}^{1/2}}
   \Y^*(-\hat{\omega},\hat{u})\partial_{\hat{u}} \Y(\hat{\omega},\hat{u}) 
+\frac{1}{\hat{u}^{1/2}}   
   \Y(-\hat{\omega},\hat{u})\partial_{\hat{u}} \Y^*(\hat{\omega},\hat{u})
      \right) +\nonumber
\\
&& y_2(-\omega)y_2(\omega)\left( \n\frac{1}{\hat{u}^{1/2}}
 \Y^*(-\hat{\omega},\hat{u})\partial_{\hat{u}} \Y(\hat{\omega},\hat{u})
 -(\n+1)\frac{1}{\hat{u}^{1/2}}
 \Y(-\hat{\omega},\hat{u})\partial_{\hat{u}} \Y^*(\hat{\omega},\hat{u})
                        \right)\Big]  \, \nonumber  \, ,
\ee
where $\n=1/(\exp\{\pi\hat{\omega}\}-1)$ and we have used the fact 
that $\y_{1,2}(\w)=y_{1,2}(\omega)$. From this expression
we can read off immediately the different correlators by taking 
derivatives with respect to the boundary values. In particular
\be
\lim_{\omega\longrightarrow 0} iG_{12}(\omega)&=&\lim_{\omega\longrightarrow 0} iG_{21}(\omega)=
\lim_{\omega\longrightarrow 0} \frac{1}{2}\left(iG_{11}(\omega)+iG_{22}(\omega)\right) \nonumber \, ,\\
&=&
\lim_{\omega\longrightarrow 0}
\frac{R^2}{\pi\alpha}\frac{2}{\pi\w} \sqrt{\gamma} 
\, {\rm Im} \left\{\frac{1}{\u^{1/2}} \Y^*(-\hat{\omega},\hat{u})\partial_{\hat{u}} \Y(\hat{\omega},\hat{u})\right\} 
\ee
Since the action for the fluctuation \Eq{action} is formally the same as that 
of the static string, we obtain the same result for the transverse momentum 
transferred as in the static case (up to an overall factor of $\sqrt{\gamma}$).
 After restoring physical units 
and using the result of \cite{Casalderrey-Solana:2006rq} and \Eq{hatqf}
\be
\label{finalq}
\kappa_T=\sqrt{\gamma \lambda} T^3 \pi \, .
\ee


\section{Conclusions}

In this paper we have studied the medium modifications of 
a heavy quark   
that propagates in a strongly 
coupled  plasma at finite velocity $v$. 
Starting from the density matrix of the 
quark, 
we have expressed the transverse momentum broadening of the probe as a 
Wilson loop running along the $x=v\,t$ line, with 
a transverse separation,  $\Delta x$. This Wilson loop is similar to
that considered in Refs.~\cite{Liu:2006he,Liu:2006ug} but approaches
the lightcone from below. The second derivative of this Wilson loop
with respect to the transverse separation $\Delta x$  yields
$\kappa_{T}$ -- the mean 
squared transverse momentum transfer per unit time acquired by
the probe in its propagation through the medium.  We have stressed that
the appropriate Wilson loop  has a time ordered line and an anti-time
ordered  line.  These two pieces can be  understood as the quark's
amplitude and complex conjugate amplitude respectively. Due to this
specific time ordering, we have introduced the type ``1" and type ``2"
fields of Schwinger-Keyldish formalism
\cite{Keldysh:1964ud,Schwinger:1960qe}.  The sources for these type ``1'' and type ``2''
fields  correspond o the boundary values of SUGRA fields in right and left quadrants of the Kruskal plane.

To construct the gravity dual of this (``1'',``2'') Wilson loop
we have extended the moving quark string solution 
found in \Refs{Yaffe,Gubser:2006bz} 
to the left Kruskal quadrant.
The string world sheet is disconnected along the $V=0$ line
because the $V=0$ line represents the distant past. In the distant past 
the moving quark string is not a solution to the equations of motion 
because the electric field used to accelerate the quark is being 
slowly turned on. 
At past infinity the two disconnected solutions are joined by
demanding analyticity in the lower half plane $V$ as 
required by the correspondence between AdS black holes and the
real time thermal field theory.

Subsequently
we have 
computed $\kappa_T$  by  
fluctuating  the transverse position of
the string endpoint and  solving for the 
standing waves. To determine the correct boundary conditions,
we  
first noticed that 
the induced metric of the string worldsheet is that of a black hole.
The event horizon of this world sheet black hole maps to
the line  $r = r_o \sqrt{\gamma}$, with $r_o$ the 
event horizon of the AdS black hole.
Then, borrowing from the work of Herzog and Son \cite{Herzog:2002pc},
we have analytically continued 
the fluctuations  through both the world-sheet and AdS horizons.
This analytic extension amounts to specifying the boundary 
conditions on the physically correct solutions in the full
Kruskal plane. 
From the solution in the full Kruskal plane 
we obtain 
\be
\kappa_T=\sqrt{\gamma \lambda} T^3 \pi \, ,
\ee
which diverges in the $v\rightarrow 1$ limit. 

Finally we wish to compare $\kappa_{T}$ (the mean squared
transverse momentum transfer  to a heavy quark per unit time)  to 
$\hat{q}$.
Up to a factor of two,
the definition of $\kappa_{T}$ given here is often ascribed to $\hat{q}$, at
least if  the qualifier ``heavy'' is removed from heavy quark.\footnote{
There is 
a trivial factor of 2 difference stemming from 
the number of spatial dimensions, 
$\kappa_{T} \rightarrow \frac{\hat{q}}{2}$\,. Further
$\hat{q}$ is often expressed in 
the adjoint representation, so that $\hat{q}_{A} = 2 \hat{q}_{F}$.
We will restrict our analysis to the fundamental representation and
the  value of $\hat{q}$ quoted below differs from the adjoint $\hat{q}$ in Eq.~(15) of Ref.~\cite{Liu:2006ug} 
by an appropriate factor of two.
}
However,
the value $2 \kappa_T$ differs from the value of the jet quenching parameter
found by Liu, Rajagopal and Wiedemann (LRW) \cite{Liu:2006he,Liu:2006ug}
\be
\hat{q}_{LRW}=\frac{\pi^{3/2}\Gamma(\frac{3}{4})}{2\Gamma(\frac{5}{4})}\sqrt{\lambda}T^3
\, .
\ee
LRW use the dipole formula as a definition
of $\hat{q}$ and compute a strictly lightlike Wilson loop. This
Wilson loop corresponds into a space-like surface. 
In this computation
special care has to be taken in approaching the limits $M\rightarrow \infty$
and $v\rightarrow 1$. Regardless of the concern expressed by some authors 
about this computation \cite{Yaffe,Chernicoff:2006hi,Argyres:2006yz} 
(most of which have been addressed in \cite{Liu:2006he}),
we do not completely understand the source of this discrepancy. 
The difference between the two results may lie in the
fact that the Wilson line in \Ref{Liu:2006he,Liu:2006ug}
is strictly light like.

Certainly the calculation described here is limited 
to the regime \footnote{This paragraph 
was triggered by discussions with  
H. Liu, K. Rajagopal and U. Wiedemann.  }
\st
\label{vbound}
          \gamma < \left(\frac{M}{\sqrt{\lambda} T} \right)^2 \, .
\stp
To see this we recall that
the quark moving with velocity $v$
was constructed by slowly turning on an electric field 
and accelerating the quark to its terminal velocity \cite{Yaffe}. The
equation of motion of the quark is  
\st
   \frac{dp}{dt} = - \eta p + E \, ,
\stp
with $\eta \sim \sqrt{\lambda} T^2/M  $.  The electric field can be increased
until its critical value, which can be computed from the Born-Infeld action for 
the probe brane in the AdS geometry\footnote{
We thank H. Liu for pointing out to us the correct value of the critical electric field.
},  
\be
S_{BI} \sim \sqrt{1-\left(2 \pi \alpha ' E \frac{R^2}{r^2}\right)^2 } \, .
\ee
The critical value for the electric field is then 
\be
E<\frac{M^2}{\sqrt{\lambda}} \,.
\ee
Equating $E \sim  \eta p$  we obtain the bound written 
above, \Eq{vbound}.  Nevertheless by taking the mass to infinity
we may approach the lightlike Wilson line from below.

In summary, the discrepancy may have its origin in the derivation of the
dipole formula or in the relation between radiative energy loss and the squared
momentum transfer. Both of these analyses 
are derived in perturbation theory. 
A careful analysis of the approximations needed to derive
these results may resolve the discrepancy 
and lead to a deeper understanding of radiation 
in a strongly coupled field theory.   \\

\noindent{\bf Note Added.}
During the completion of this work a preprint by S. Gubser,  hep-th/0612143, 
appeared and revealed the properties of the  world sheet 
black hole.  We gratefully acknowledge illuminating 
discussions over the past weeks with Professor
Gubser, who pointed out (amongst other things) an algebraic
error in our draft which had led to a $\kappa_T$ independent
of the quark velocity. Even after correcting this error
this manuscript did not agree with the first version of hep-th/0612143.
The differences have since been resolved.


\noindent {\bf Acknowledgments.} 
J.C thanks J. L. Albacete, N. Armesto, 
V. Koch,
C. Salgado, X. N. Wang, U. A. Wiedemann
for useful discussions about the computation of 
radiative losses and  H. Liu about the computation
of the light-like Wilson Loops.
This work was supported by the Director, 
Office of Science, Office of High Energy and Nuclear Physics, 
Division of Nuclear Physics, and by the Office of Basic Energy
Sciences, Division of Nuclear Sciences, of the U.S. Department of Energy 
under Contract No. DE-AC03-76SF00098.

\end{document}